\documentclass[amsmath,amssymb,aps,prl, twocolumn,superscriptaddress,notitlepage]{revtex4-2}

\usepackage{graphicx}
\usepackage{bbm}

\usepackage{hyperref}
\usepackage{xcolor}
\begin{document}
\preprint{APS/123-QED}
\newcommand{\pto}{ PbTi$\text{O}_3$ }
\newcommand{\bto}{ BaTi$\text{O}_3$ }
\newcommand{\red}[1]{\textcolor{red}{#1}}
\newcommand{\blue}[1]{\textcolor{blue}{#1}}

\title{Ab Initio bulk free energy surface of proper ferroelectrics}
 
\author{Pinchen Xie}
\affiliation{Applied Mathematics and Computational Research Division, Lawrence Berkeley National Laboratory, Berkeley, CA 94720, USA}
\affiliation {Program in Applied and Computational Mathematics, Princeton University, Princeton, NJ 08544, USA}

\author{Yixiao Chen}
\affiliation {Program in Applied and Computational Mathematics, Princeton University, Princeton, NJ 08544, USA}

\author{Xinyu Xu}
\affiliation{Department of Materials Science and Engineering, Stanford University, Stanford, CA 94305, USA}

\author{Zhi Yao}
\affiliation{Applied Mathematics and Computational Research Division, Lawrence Berkeley National Laboratory, Berkeley, CA 94720, USA}

\author{Weinan E}
\thanks{Current address: AI for Science Institute, Beijing, China, Center for Machine Learning Research and School of Mathematical Sciences, Peking University, Beijing, China}
\affiliation {Program in Applied and Computational Mathematics, Princeton University, Princeton, NJ 08544, USA}

\author{Roberto Car}
\email{rcar@princeton.edu}
\affiliation {Department of Chemistry, Department of Physics, Princeton Materials Institute, Princeton University, Princeton, NJ 08544, USA}
\affiliation {Program in Applied and Computational Mathematics, Princeton University, Princeton, NJ 08544, USA}
\date{\today}
\begin{abstract}
We report a systematic and accurate approach for deriving the bulk free energy surface (FES), a function of temperature, polarization, and strain, from the first-principles density functional theory (DFT) of proper ferroelectrics. The core of our approach is the metadynamics algorithm that extracts the polarization dependence of the FES from all-atom molecular dynamics simulations without an {\it a priori} ansatz. The rest of the FES is derived from the metadynamics trajectories that span the relevant phase space. We demonstrate our approach in the case of lead titanate. The errors across the phase transition, due to DFT numerics, all-atom molecular dynamics, and free energy evaluation by enhanced sampling, can be systematically controlled and are of the order of 1meV/atom. The accuracy of the resulting {\it ab initio} FES is only limited by the adopted functional approximation of DFT. 
\end{abstract}

\maketitle

The bulk free energy surface (FES) $\mathcal{F}(T,\boldsymbol{\mathcal{P}}, \boldsymbol{\eta})$, as a function of temperature ($T$), polarization vector ($\boldsymbol{\mathcal{P}}$) and global strain tensor ($\boldsymbol{\eta}$), characterizes the thermodynamic, dielectric and structural properties of proper ferroelectric materials~\cite{chandra2007landau}. 
It is the major component of the time-dependent Ginzburg-Landau (TDGL) models~\cite{chen2002phase}, representing the mean-field energy density of a homogeneous region in isothermal and isobaric equilibrium.
TDGL models further include terms depending on the gradient of the order parameter field to describe macroscopic inhomogeneities induced by interfaces or grain boundaries originating smooth near-equilibrium variations of the order parameter.
Such models can predict mesoscopic structural evolution~\cite{xue2021theory, yadav2016observation}  and device metrics such as capacitance and I-V response~\cite{kumar2024,kumar2023ferrox,saha2020,yadav2019spatially} beyond the reach of atomistic models. Therefore, accurate deriving of the bulk FES from first principles is desirable for material discovery, characterization, and  device engineering. 

The phenomenological Landau-Devonshire (LD) theory~\cite{devonshire1949xcvi,devonshire1951cix,devonshire1954theory} approximates the bulk FES with a polynomial of $T$, $\boldsymbol{\mathcal{P}}$ and $\boldsymbol{\eta}$ with parameters fitted to the experiments.  The bulk FES underlying the phenomenological models is defined microscopically by $\mathcal{F}(T,\boldsymbol{\mathcal{P}}, \boldsymbol{\eta}) = -k_BT \ln  \int d \mathbf{R} e^{-\beta \mathcal{H}(\mathbf{R})}   \delta(\boldsymbol{\mathcal{P}}(\mathbf{R}) - \boldsymbol{\mathcal{P}})\delta(\boldsymbol{\eta}(\mathbf{R}) - \boldsymbol{\eta})$, 
in terms of the adiabatic potential energy surface  $\mathcal{H}(\mathbf{R})$, the polarization surface $\boldsymbol{\mathcal{P}}(\mathbf{R})$, and the strain tensor surface $\boldsymbol{\eta}(\mathbf{R})$ associated with an atomic configuration $\mathbf{R}$. Here, $\beta=1/k_BT$, $\delta$ is the Dirac delta function, $\boldsymbol{\mathcal{P}}(\mathbf{R})$ is prescribed by the theory of polarization~\cite{vanderbilt2018berry, resta2007theory} with a gauge redundancy \footnote{The redundancy is eliminated when a reference structure is chosen for the zero of polarization.} and a periodic boundary condition \footnote{ The periodicity of the polarization is associated with the polarization quantum~\cite{vanderbilt2018berry, resta2007theory}. In practice, the change of polarization typically does not exceed the quantum of polarization. Thus, polarization can often be treated as a single-value function, as we do in this paper}.
 
Ab initio density functional theory (DFT) has provided access to both $\mathcal{H}(\mathbf{R})$ and $\boldsymbol{\mathcal{P}}(\mathbf{R})$. But ab initio calculation of the bulk FES was obstructed by two major difficulties. The first is the high computational cost of DFT calculations, an issue that has recently been overcome by neural network models that parameterize $\mathcal{H}(\mathbf{R})$ and $\boldsymbol{\mathcal{P}}(\mathbf{R})$ with DFT precision~\cite{PhysRevLett.120.143001, zhang2020dielectric}. The second is that direct molecular dynamics (MD) or Monte Carlo (MC) simulations driven by
$\mathcal{H}(\mathbf{R})$ are often insufficient to sample rare events associated with crossing large energy barriers between different phases. 
Neural network-accelerated MD simulations cannot fully address this issue because the probability of barrier crossing decreases exponentially with barrier height. Enhanced sampling methods~\cite{barducci2011metadynamics, valsson2014variational, PhysRevX.10.041034} have been developed to overcome this difficulty. While these methods are widely used to study chemical reactions and molecular conformational changes, they are often limited to one or two dimensional (1D/2D) order parameters when a highly accurate FES (with an error margin within $k_BT$) is required. 
In the ferroelectric problem, the polarization is 3D and the strain adds another three degrees of freedom, if shear strain is not present, and six degrees of freedom, in the general case.   
 
\begin{figure}[t]
    \centering
    \includegraphics[width=\linewidth]{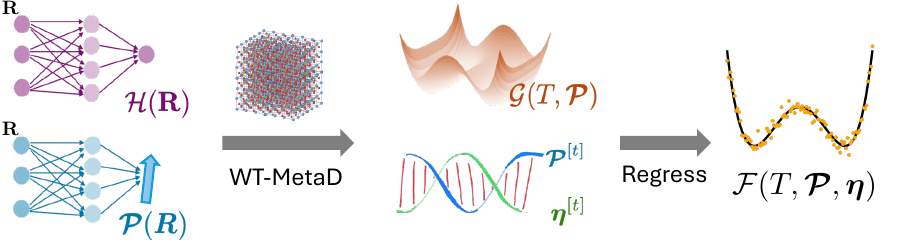}
    \caption{Schematic representation of the approach}
    \label{fig:workflow}
\end{figure}

This work aims to overcome the difficulty posed by dimensionality and develop a systematic approach to extract the bulk FES of proper ferroelectric crystals, with potential extension to other proper ferroic materials, such as magnetic crystals, ferroelastic crystals, and their antiferroic counterparts. Our approach includes the three steps illustrated in Fig.~\ref{fig:workflow}. The first step consists of training neural network models for $\mathcal{H}(\mathbf{R})$ and $\boldsymbol{\mathcal{P}}(\mathbf{R})$ on DFT data. Linear polarization changes with the atomic displacements could be described by Born charge models, but here we consider neural network models for $\boldsymbol{\mathcal{P}}(\mathbf{R})$, whose validity extends beyond the linear response for the atomic displacements.   
The second step uses well-tempered metadynamics simulations (WT-MetaD~\cite{PhysRevLett.100.020603}) with flexible cells and regularized by symmetry to calculate the marginal distribution $e^{-\beta \mathcal{G}(T,\boldsymbol{\mathcal{P}})} \propto \int_{\boldsymbol{\eta}} e^{-\beta \mathcal{F}(T,\boldsymbol{\mathcal{P}},\boldsymbol{\eta})} d\boldsymbol{\eta}$ in the relevant regime of $T$ and $\boldsymbol{\mathcal{P}}$. WT-MetaD does not require an {\it a priori} ansatz for the $\boldsymbol{\mathcal{P}}$ dependence of $\mathcal{G}(T,\boldsymbol{\mathcal{P}})$. However, the last step of the approach is based on an {\it a priori} ansatz from quadratic electrostrictive theory for the $\boldsymbol{\eta}$ dependence of $\mathcal{F}(T,\boldsymbol{\mathcal{P}},\boldsymbol{\eta}) - \mathcal{G}(T,\boldsymbol{\mathcal{P}})$, considering the higher-order correction to strain-dependence in $\mathcal{F}$ is insignificant under moderate external electric fields, pressures, and epitaxial strains. The validity of the ansatz will be demonstrated by analyzing the correlated fluctuations of the polarization and cell parameters in WT-MetaD simulations. 

The enhanced sampling algorithm WT-MetaD lies at the core of our approach. Within the framework of isobaric-isothermal MD, WT-MetaD adds a time-dependent bias potential $V^{[t]}(\boldsymbol{\lambda}(\mathbf{R}))$ to the microscopic potential $\mathcal{H}(\mathbf{R})$. Here, $\boldsymbol{\lambda}(\mathbf{R})$ is a multidimensional order parameter, differentiable in $\mathbf{R}$. $V^{[t]}$ depends on the previous history of the system and varies slowly on the time scale of a simulation; it applies to the system a biasing force $-\frac{\partial V^{[t]}(\boldsymbol{\lambda}(\mathbf{R}))}{\partial \mathbf{R}}$, which is calculated on the fly, to explore the relevant domain of the order parameter.
In this work, $\boldsymbol{\lambda}(\mathbf{R})$ is the polarization vector $\boldsymbol{\mathcal{P}}(\mathbf{R})$ described by a neural network model that allows efficient gradient calculations.
Through NPT-MD simulations, $V^{[t]}$ is adaptively modified so that asymptotically it approaches $(\gamma^{-1}-1)\mathcal{G}(T,\boldsymbol{\mathcal{P}})$, where the value of the scalar parameter $\gamma$ is set to facilitate frequent crossings between the paraelectric and ferroelectric basins of the free-energy landscape
(see Appendix B for details). Using WT-MetaD,
the asymptotic bias potential, and hence the FES, is represented directly on a dense grid of the order parameter, without needing a parameterized interpolation ansatz. 

The generality of WT-MetaD distinguishes our study from previous works in which $\mathcal{G}(T,\boldsymbol{\mathcal{P}})$ was recovered from microscopic simulations by assuming an analytic form {\it a priori}~\cite{iniguez2001devonshire}, or by calculating free energy differences along 1D thermodynamic paths~\cite{ Geneste2009, Kumar2010}. 
Moreover, to our knowledge, no previous work provided a systematic evaluation of $\mathcal{F}(T,\boldsymbol{\mathcal{P}},\boldsymbol{\eta})$, including strain effects.
Here, we provide not only the temperature-dependent values of  parameters that define $\mathcal{F}(T,\boldsymbol{\mathcal{P}},\boldsymbol{\eta})$ but also the zero-strain cell parameters that underlie the definition of $\boldsymbol{\eta}$. 

\textit{Results} - 
We demonstrate our approach in the test case of PbTi$\text{O}_3$, a prototypical ferroelectric material for which the sixth-order LD model should describe well the dependence of the FES on polarization in the vicinity
of the ferroelectric-paraelectric (FE-PE) phase transition. We use a Deep Potential (DP) neural network model~\cite{PhysRevLett.120.143001, zhang2018end}, and a Deep Dipole (DD)~\cite{zhang2020dielectric} neural network model to represent, respectively, $\mathcal{H}(\mathbf{R})$, and $\boldsymbol{\mathcal{P}}(\mathbf{R})$. The two models are trained on DFT data generated with the Strongly Constrained and Appropriately Normed (SCAN) functional approximation~\cite{sun2015strongly} using an active learning strategy~\cite{zhang2019active,ZHANG2020107206}. In a representative set of configurations independent of the training data, the root mean square error of the DP model was $1.0$ meV per atom for energy prediction and $0.35\mathrm{eV/\AA}$ per atom for force prediction. The root mean square error of the DD model for polarization prediction was $1.4\mathrm{\mu C/cm^2}$. The gauge redundancy of $\boldsymbol{\mathcal{P}}$ was fixed by setting the polarization of the centrosymmetric perovskite structure to zero. 
The details of these two models are provided in a previous publication~\cite{xie2024pto}, where an extensive molecular dynamics study of the FE-PE transition was reported, showing good agreement with experimental results on structural, dielectric, and spectroscopic properties. 

\begin{figure}[t]
    \centering
    \includegraphics[width=0.9\linewidth]{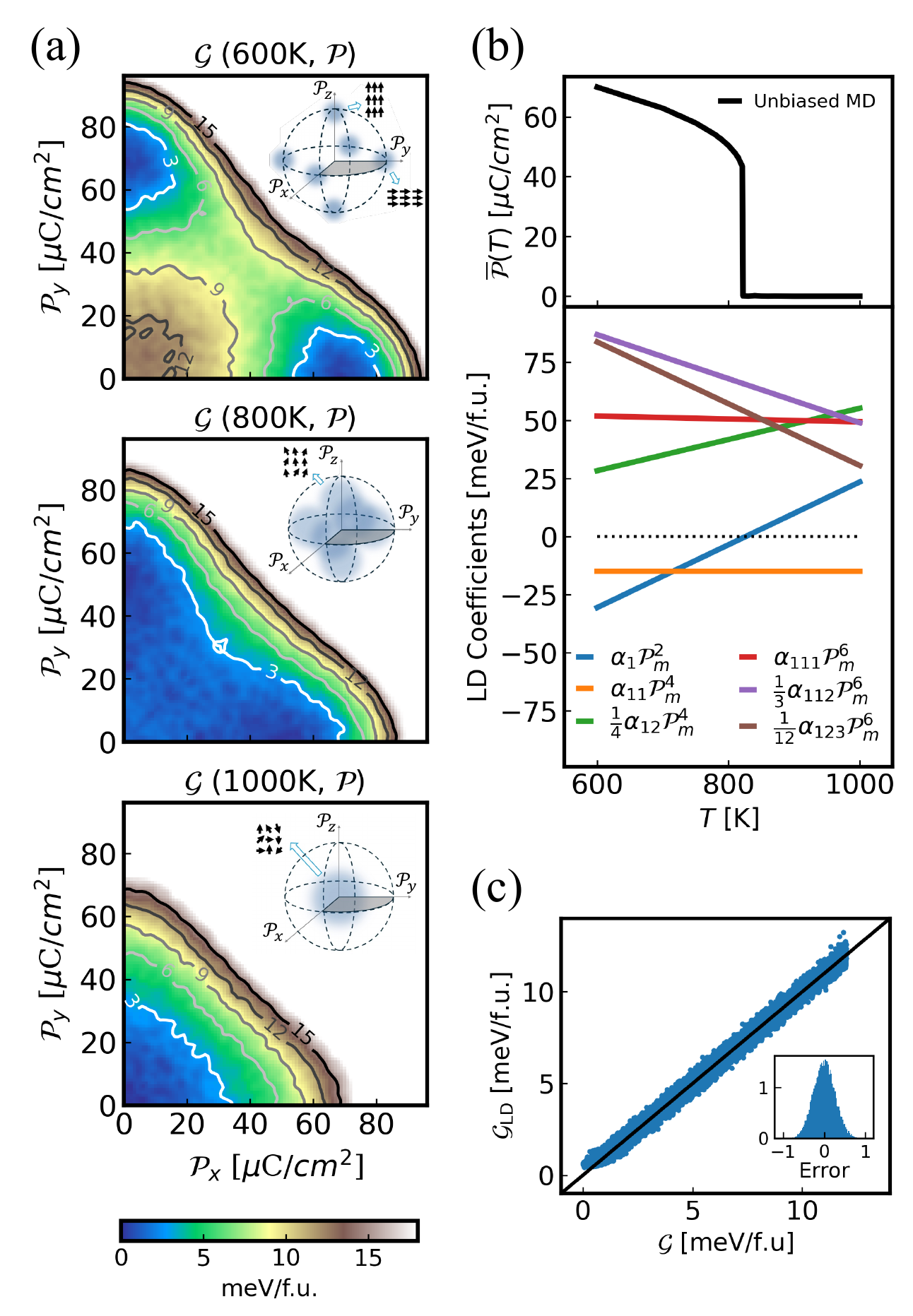}
    \caption{
    (a) Colored cross-sectional ($\mathcal{P}_z=0$) representation of $\mathcal{G} (T, \boldsymbol{\mathcal{P}})$. The contour lines are equipotential lines.
   Each panel in inset contains a sketch of the energy basins in $\mathcal{G} (T, \boldsymbol{\mathcal{P}})$ as blue shades. The arrows illustrate local dipole configurations. 
    (b) Top: spontaneous polarization from unbiased MD simulations~\cite{xie2024pto}.
    Bottom: Optimal LD coefficients (scaled) as functions of temperature. Scaling factors are indicated in the legends. 
    (c) Scatter plot of all the $\mathcal{G} (T, \boldsymbol{\mathcal{P}})$ data versus $\mathcal{G}_{\text{LD}}(T, \boldsymbol{\mathcal{P}})$. The inset shows the error distribution. }
    \label{fig:fes2D}
\end{figure}

A critical finding of~\cite{xie2024pto} was that the isobaric FE-PE transition in PbTi$\text{O}_3$ is driven by dipolar fluctuations occuring homogeneously at the nanoscale, rather than by a nucleation and growth mechanism like the one that drives the ferroelectric domain switching in this material~\footnote{A direct molecular dynamics simulation of the ferroelectric domain switching (driven by nucleation and growth) in PbTi$\text{O}_3$ with our Deep Potential model can be found in Sec. 3.1 of ~\cite{xie2024thesis}}. Classical nucleation theory does not apply because of the presence of strong elastic-dipolar coupling and significant lattice mismatch between the PE and FE states. The elastic strain generated by lattice mismatch at a coherent PE-FE interface is long-ranged and induces a volume law energy cost~\cite{cahn1984simple, schwarz1995thermodynamics, jin2019macroscopic} at the submicrometer scale, instead of the area-law cost associated with incoherent interfaces such as those between coexisting solid and liquid phases. Then, a homogenous transition is favored over nucleation and growth~\cite{xie2024pto}. Correspondingly, the bulk FES per formula unit (f.u.) is a size-independent mean-field energy density that can be accurately estimated with finite-size WT-MetaD simulations. 
In Appendix A, we report a finite-size analysis of the 1D free energy profile $G(T, d_c)=-k_BT \ln\int d\mathbf{R} e^{-\beta \mathcal{H}(\mathbf{R})} \delta(\|\boldsymbol{\mathcal{P}}(\mathbf{R})\Omega(\mathbf{R})\| - d_c)$, where $\Omega(\mathbf{R})$ and $d_c$ are the average volume and dipole moment magnitude of the primitive cell. The analysis confirms the volume-law scaling of the free energy barrier that separates the PE and FE states. It further indicates that, for our DP model, the phase transition temperature $T_c=(821\pm 1)$ K, in good agreement with the experimental value of $T_c^{\mathrm{exp}}=763$K. In addition, it suggests that the finite-size error in the FES should be of the order of 0.1meV per formula unit (f.u.) when $L\geq 9$ for flexible perovskite supercells containing $5L^3$ atoms. We adopt $L=9$ in all the simulations reported below. 

Next, as illustrated in Fig.~\ref{fig:workflow}, we compute $\mathcal{G}(T,\boldsymbol{\mathcal{P}})$ with WT-MetaD (see Appendix B for details). Since $\mathcal{G} (T, \boldsymbol{\mathcal{P}})$ is invariant under mirror reflections and permutations of the Cartesian components of $\boldsymbol{\mathcal{P}}=(\mathcal{P}_x, \mathcal{P}_y, \mathcal{P}_z)$, we restrict the WT-MetaD sampling domain to the sector $\mathcal{P}_{m} \geqslant \mathcal{P}_{x} \geqslant \mathcal{P}_{y} \geqslant \mathcal{P}_{z} \geqslant 0$~\footnote{The restriction can be applied as restraining potentials ($\text{UPPER\_WALLS}$ and $\text{LOWER\_WALLS}$) in PLUMED. Relevant scripts are provided in Supplementary Code and Data~\cite{alldata}}. We use a cut-off value $\mathcal{P}_{m}= 96 \mu \text{C}/\text{cm}^2$, adopt a dense 3-D grid to represent the $\boldsymbol{\mathcal{P}}$ sector, and compute $\mathcal{G} (T, \boldsymbol{\mathcal{P}})$ on the dense grid at discrete temperature values ($T\in[600,700,800,820,900,1000]$K). At each $T$, we run an 8 ns-long WT-MetaD trajectory, sufficient for good convergence. Enhanced sampling methods like WT-MetaD yield free energy differences, not absolute free energies.
Thus, at each temperature $T$, the calculated $\mathcal{G} (T, \boldsymbol{\mathcal{P}})$ is only known modulo an arbitrary constant, which is fixed here by requiring that $\min_{\boldsymbol{\mathcal{P}}} \mathcal{G} (T, \boldsymbol{\mathcal{P}})=0$ . The cross sections of the computed $\mathcal{G} (T, \boldsymbol{\mathcal{P}})$ for $\mathcal{P}_z=0$ are plotted in Fig.~\ref{fig:fes2D}(a) for three representative temperatures. With increasing temperature, from 600K to 1000K, the thermal disorder increases and eliminates long-range ferroelectric order~\cite{xie2024pto}, leading to a stable paraelectric basin. The top panel of Fig.~\ref{fig:fes2D}(b) shows the corresponding change in spontaneous thermodynamic polarization, obtained from unbiased MD simulations~\cite{xie2024pto}. 
The sketches in the upper right corner of each panel in Fig.~\ref{fig:fes2D}(a) provide a schematic representation of the evolution of the FES. 

$\mathcal{G} (T, \boldsymbol{\mathcal{P}})$ extracted from the WT-MetaD simulations is affected
by stochastic noise of the order of $0.2\sim 0.3$ meV/f.u. at each point in the grid. One can use a posterior ansatz to denoise and interpolate the results over the entire $T\in [600,1000]$K range. The posterior ansatz should be able to fit all WT-MetaD results with controlled error. Here we use the sixth-order LD ansatz with temperature-dependent coefficients, given by  
\begin{equation}\label{landau}
\small
    \begin{split}
    &\mathcal{G}_{\text{LD}}=G_0 + \alpha_1\|\boldsymbol{\mathcal{P}}\|^2 
    +\alpha_{11}(\mathcal{P}_x^4+\mathcal{P}_y^4+\mathcal{P}_z^4)
    +\alpha_{12}(\mathcal{P}_x^2\mathcal{P}_y^2 \\
    &+\mathcal{P}_x^2\mathcal{P}_z^2 +\mathcal{P}_y^2\mathcal{P}_z^2  )
    +\alpha_{111}(\mathcal{P}_x^6+\mathcal{P}_y^6+\mathcal{P}_z^6)+\alpha_{123}\mathcal{P}_x^2\mathcal{P}_y^2\mathcal{P}_z^2\\
    &+\alpha_{112}[
        \mathcal{P}_x^4(\mathcal{P}_y^2+\mathcal{P}_z^2)
        +\mathcal{P}_y^4(\mathcal{P}_z^2+\mathcal{P}_x^2)
        +\mathcal{P}_z^4(\mathcal{P}_x^2+\mathcal{P}_y^2)].
    \end{split}
\end{equation}

All LD coefficients, except immaterial $G_0$, are represented by linear functions of $T$. $G_0=G_0(T)$ merely shifts the minima of $\mathcal{G}_{\text{LD}}$ to zero and is set separately for each temperature. We fit the LD ansatz to all the $\mathcal{G} (T, \boldsymbol{\mathcal{P}})$ data extracted from the WT-MetaD trajectories in the range $\mathcal{G}\leq 12$ meV / f.u., for which the simulations provide good statistics.
The least squares fit (see Supplemental Material for details) leads to the optimal LD coefficients that are reported in the bottom panel of Fig.~\ref{fig:fes2D}(b). The optimal $\alpha_1(T)$ yields a Curie temperature $T_\theta=826$K and a Curie constant $C\approx 1.4\times 10^5$K close to the value $C^{\mathrm{EXP}}\approx 1.5\times 10^5$K~\cite{haun1987thermodynamic} extracted from measurements in powder samples~\cite{haun1987thermodynamic}. 
In the same experiment, values of the other LD parameters, assumed to be temperature independent, were estimated, which, when converted to our convention and units (meV/f.u.) read $\alpha_{11}^{\text{EXP}}\mathcal{P}_m^2=-24$, $\frac{1}{4}\alpha_{12}^{\text{EXP}}\mathcal{P}_m^2=60$, $\alpha_{111}^{\text{EXP}}\mathcal{P}_m^6=80$, $\frac{1}{3}\alpha_{112}^{\text{EXP}}\mathcal{P}_m^6=63$ and $\frac{1}{12}\alpha_{123}^{\text{EXP}}\mathcal{P}_m^6=-95$. All phenomenological parameters, with the exception of $\alpha_{123}^{\text{EXP}}$, are in semiquantitative agreement with our results. The difference between powder samples and the ideal single crystal suggests that the experimental high-order parameters could deviate the most from the intrinsic material properties. This may explain why only the lowest-order parameter $\alpha^{\text{EXP}}_1$ agrees closely with our results. 



A correlation plot of $\mathcal{G}_{\text{LD}} (T, \boldsymbol{\mathcal{P}})$ versus $\mathcal{G} (T, \boldsymbol{\mathcal{P}})$ extracted from WT-MetaD in the sampled range of temperatures and polarizations is reported in Fig.~\ref{fig:fes2D}(c). 
The fitting residue $\mathcal{G}_{\text{LD}}- \mathcal{G}$ is normally distributed (see inset of Fig.~\ref{fig:fes2D}(c)) with a standard deviation of $0.26$ meV/f.u. The standard deviation matches the stochastic error ($0.2\sim 0.3$ meV/f.u.) in the numerical $\mathcal{G} (T, \boldsymbol{\mathcal{P}})$ on the dense $\boldsymbol{\mathcal{P}}$ grid, suggesting that the optimal $\mathcal{G}_{\text{LD}}(T, \boldsymbol{\mathcal{P}})$ has denoised the WT-MetaD results with the same level of accuracy of the underlying SCAN-DFT approach (see Appendix B for visual comparisons of heatmaps of $\mathcal{G} (T, \boldsymbol{\mathcal{P}})$ and $\mathcal{G}_{\text{LD}} (T, \boldsymbol{\mathcal{P}})$). 
In the above, we considered the LD theory as a mere regression tool. For systems unlike standard perovskites that may require LD models beyond 6th order, one could always adopt a more complex posterior ansatz to denoise and interpolate the WT-MetaD data in the desired range of interest.


Next, we extract the strain dependence of $\mathcal{F}(T,\boldsymbol{\mathcal{P}}, \boldsymbol{\eta})$ from the fluctuations of the simulation box in the WT-MetaD trajectories. We restricted the simulation to anisotropic, orthogonal cell fluctuations due to the lack of rhombohedral phases in \pto. Thus, only the three diagonal components of the strain tensor are monitored, that is, $\boldsymbol{\eta}= (\eta_x, \eta_y, \eta_z)\equiv (\eta_{xx}, \eta_{yy}, \eta_{zz})$ in the Voigt notation. For a supercell with instantaneous cell parameters $\boldsymbol{l}=(l_x,l_y,l_z)$, we define the instantaneous strain components to be $\eta_i = \frac{l_i-l^{\text{ref}}_i(T)}{l^{\text{ref}}_i(T)}$ for $i=x,y,z$. The reference cell parameters $l^{\text{ref}}_i(T)$ are extracted from the WT-MetaD trajectories as explained below. 
 
Within the theory of quadratic electrostriction~\cite{chen2007appendix}, the strain dependence in $\mathcal{F}(T,\boldsymbol{\mathcal{P}}, \boldsymbol{\eta})$ can be attributed to two contributions, the elastic energy $E_{\text{elas}}(T,\boldsymbol{\eta}) = \frac{N}{2}  \boldsymbol{\eta}^T \hat{B} \boldsymbol{\eta}$ and the strain-polarization coupling energy $E_{\text{coup}}(T, \boldsymbol{\mathcal{P}},\boldsymbol{\eta}) = -N \boldsymbol{Q}^T \hat{\Gamma}  \boldsymbol{\eta}$, where $N=L^3$ is the number of formula units in the simulation. The $3\times 3$ matrices $\hat{B}$ and $\hat{\Gamma}$ are temperature-dependent elastic and electrostrictive constant matrices, respectively. $\boldsymbol{Q}^T$ is the row vector $(\mathcal{P}_x^2, \mathcal{P}_y^2,\mathcal{P}_z^2)$.  For simplicity, we focus on the isotropic case, for which $l_x^{\text{ref}}(T) = l_y^{\text{ref}}(T) =l_z^{\text{ref}}(T) \equiv l^{\text{ref}}(T)$ and the matrix
$\hat{B}$ has equal diagonal elements denoted by $B_{11}(T)$ and equal off-diagonal elements denoted by $B_{12}(T)$. The same holds for $\hat{\Gamma}$. 
 
The statistical mechanics consistency condition $e^{-\beta \mathcal{G}(T,\boldsymbol{\mathcal{P}})} \propto \int_{\boldsymbol{\eta}} e^{-\beta \mathcal{F}(T,\boldsymbol{\mathcal{P}},\boldsymbol{\eta})} d\boldsymbol{\eta}$ requires that $\mathcal{F}$ be related to $\mathcal{G}$ through
$\mathcal{F}  =\mathcal{G}  +  \frac{N}{2} (\boldsymbol{\eta}^T-\boldsymbol{Q}^T\hat{Z}^T) \hat{B} (\boldsymbol{\eta} - \hat{Z}\boldsymbol{Q})$
with $\hat{Z}\equiv \hat{B}^{-1} \hat{\Gamma}$. In our case $\hat{Z}$ has equal diagonal elements ($Z_{11}(T)$) and equal off-diagonal elements ($Z_{12}(T)$).
For given $T$ and $\boldsymbol{\mathcal{P}}$, the bulk FES is minimal when $\boldsymbol{\eta}=\hat{Z}\boldsymbol{Q}$. The correlation of the fluctuations of the residual vector $\boldsymbol{\zeta}=\boldsymbol{\eta}-\hat{Z}\boldsymbol{Q}$ is given by the covariance matrix $(\beta N)^{-1}\hat{B}^{-1}$. Since, for a sufficiently large $N$, the fluctuations of $\boldsymbol{\zeta}$ should follow a 3D normal distribution, we can determine $l^{\text{ref}}$, $\hat{B}$, and $\hat{\Gamma}$ from the trajectories of $\boldsymbol{l}$ and $\boldsymbol{\mathcal{P}}$ using the standard theory of multivariate normal distributions. Specifically, optimal $l^{\text{ref}}$ and $\hat{Z}$ are extracted at each temperature from the least squares fit that minimizes $\|\boldsymbol{\zeta}\|^2$, and $\hat{B}$ is obtained by inverting the covariance matrix of optimal $\boldsymbol{\zeta}$ (see Supplemental Material for details). Trajectories from WT-MetaD, rather than from unbiased MD, are necessary because $\|\boldsymbol{\zeta}\|^2$ should be minimal in the entire phase space domain of interest, rather than only in a small basin near a FES minimum.

\begin{figure}[t]
    \centering
    \includegraphics[width=0.9\linewidth]{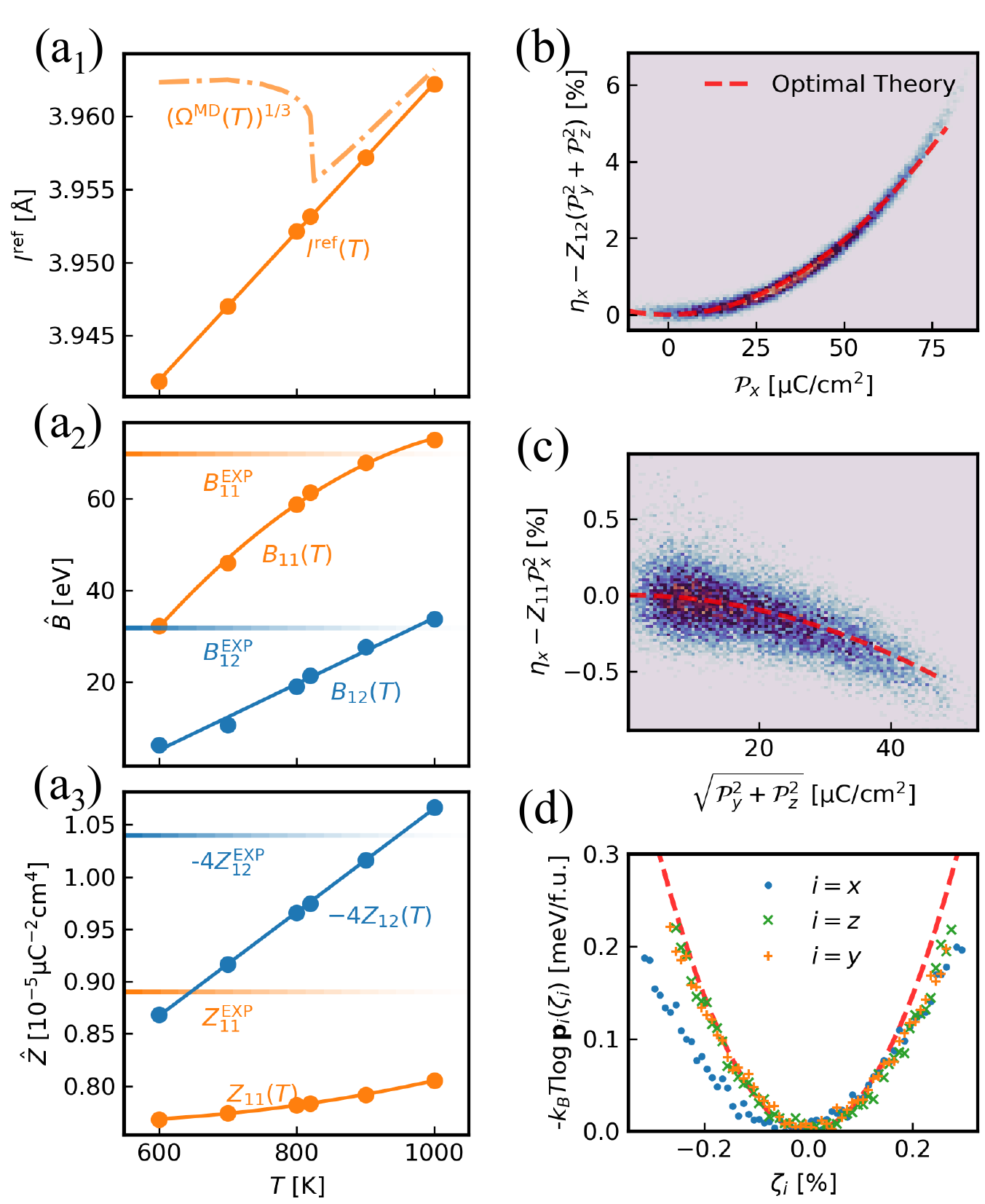}
    \caption{
    ($\mathrm{a_1}$-$\mathrm{a_3}$) Optimal $l^{\text{ref}}$, $\hat{B}$, and $\hat{Z}$ as functions of temperature. Circles are from direct least squares fits. Solid lines are their linear or quadratic interpolation. Fading lines are phenomenological estimates assuming $T$ independence from Ref.~\cite{haun1987thermodynamic}.  
    (b-c) 2D histograms of $\mathrm{p}(\mathcal{P}_x, \eta_x - Z_{12}(\mathcal{P}^2_y+\mathcal{P}^2_z))$ and $\mathrm{p}(\sqrt{\mathcal{P}_y^2+\mathcal{P}_z^2}, \eta_x - Z_{11}Q_x)$, obtained at $T=800$K. Red lines represent $\boldsymbol{\eta}=\hat{Z}(800\text{K})\boldsymbol{Q}$. 
    (d) Comparison between $- k_BT \log \mathrm{p}_i(\zeta_i)$ (shifted by $N\epsilon_{0,i}(T)$ and represented by blue, green and orange markers) and the optimal theory $\frac{1}{2} N \zeta_i^2 B_{11}(T)$ (red dashed line). }
    \label{fig:fes-strain}
\end{figure}

The optimal parameters obtained in this way at different temperatures are shown in Fig.~\ref{fig:fes-strain}($\mathrm{a_1}$-$\mathrm{a_3}$). 
$l^{\text{ref}}(T)$, $B_{12}(T)$ and $Z_{12}(T)$ exhibit excellent linearity with temperature. The corresponding linear interpolations are plotted as solid lines. The optimal $B_{11}(T)$ and $Z_{11}(T)$ are accurately interpolated by quadratic polynomials, also shown as solid lines in the figure. It is interesting to compare $l^{\text{ref}}(T)$ with the cubic root of the average volume per formula unit, $(\Omega^{\text{MD}}(T))^{1/3}$, extracted from unbiased MD. While $l^{\text{ref}}(T)$ shows perfect linear behavior that extends from the cubic phase for $T>T_c$ to the tetragonal phase for $T<T_c$, $(\Omega^{\text{MD}})^{1/3}$ shows a sharp discontinuity at $T_c$, as expected for a first-order phase transition. In fact, it would be wrong to identify an observable like $(\Omega^{\text{MD}})^{1/3}$ with a Landau parameter like the ``strain-free'' cell parameter $l^{\text{ref}}(T)$, which is a continuous function of $T$ throughout the phase transition, like the LD parameters in Eq.~$\eqref{landau}$. Observables are associated with the minima of the Landau FES which change discontinuously across $T_c$. In the present case, the dependence of the FES on $\boldsymbol{\eta}$ is quadratic, but the coupling of $\boldsymbol{\eta}$ with $\boldsymbol{\mathcal{P}}$ leads to discontinuities in observables that depend on $\boldsymbol{\eta}$ like $\Omega$.

The optimal matrices $\hat{B}(T)$ and $\hat{Z}(T)$
predicted here can be compared with experimental estimates that assume temperature independence. The values extracted in this way from powder samples ~\cite{haun1987thermodynamic} are reported
in Fig.~\ref{fig:fes-strain} 
($\mathrm{a_2}$-$\mathrm{a_3}$)) as fading lines that are typically in the range of our theoretical predictions.     
To further validate the optimal fitting, we plot
in Figs.~\ref{fig:fes-strain}(b-c) the 2D probability density distributions $\mathrm{p}(\mathcal{P}_x, \eta_x - Z_{12}(\mathcal{P}_y^2+\mathcal{P}_z^2))$ and $\mathrm{p}(\sqrt{\mathcal{P}_y^2+\mathcal{P}_z^2}, \eta_x - Z_{11}\mathcal{P}_x^2)$ extracted from the WT-MetaD simulations at $T=800$K. These figures indicate that the maximum intensity of both distributions almost coincides with
the optimal strain tensor $\boldsymbol{\eta}=\hat{Z}\boldsymbol{Q}$ shown as a dashed red line. 
The fitting error can be more rigorously quantified by $\epsilon(T,\boldsymbol{\zeta}) = - N^{-1}k_BT \log \mathrm{p}(\boldsymbol{\zeta})- \frac{1}{2} \boldsymbol{\zeta}^T \hat{B}(T) \boldsymbol{\zeta}  + \epsilon_0(T)$, where $\epsilon_0(T)$ is an immaterial constant that shifts $\epsilon(T,0)$ to zero. Here, $\mathrm{p}(\boldsymbol{\zeta})$ is the 3D probability density distribution of the residual. $\epsilon(T, \boldsymbol{\zeta})$ converges slowly with time along the WT-MetaD trajectories, and it is
easier to compute the logarithmic likelihood distances between the marginal distributions of $\mathrm{p}(\boldsymbol{\zeta})$ and the optimal theory, denoted by $\epsilon_i(T,\zeta_i) = - N^{-1}k_BT \log \mathrm{p}_i(\zeta_i)- \frac{1}{2} \zeta_i^2 B_{11}(T) + \epsilon_{0,i}(T)$. Here, $\mathrm{p}_i(\zeta_i)=\int\mathrm{p}(\boldsymbol{\zeta})d\zeta_j d\zeta_k$. $(i,j,k)$ is any permutation of $(x,y,z)$. Fig.~\ref{fig:fes-strain}(d) compares $- N^{-1}k_BT \log \mathrm{p}_i(\zeta_i)+\epsilon_{0,i}(T)$ with $\frac{1}{2} \zeta_i^2 B_{11}(T)$ at $T=800$K. For $i=y$ and $i=z$, the error is on the order of $0.01$meV/f.u., while $\epsilon_x(T,\zeta_x)$, which is associated with the direction of spontaneous polarization in the simulations, exhibits a larger error on the order of $0.1$meV/f.u.. The same trends and orders-of-magnitude errors are observed at all temperatures. The small but systematic deviations should be attributed to higher-order couplings between $\boldsymbol{\eta}$ and $\boldsymbol{\mathcal{P}}$ that are ignored here but become appreciable when $\boldsymbol{\mathcal{P}}$ is large. It seems unnecessary to correct for such deviations by replacing quadratic with quartic theory because the error is of the same order as the statistical error of the WT-MetaD data.

\textit{Discussion} - We have introduced a systematic coarse-graining approach to bridge DFT, MD, and the bulk FES of proper ferroelectric materials with controlled errors at all stages of the procedure.
The error of the PES for MD simulations is of the order of $1$meV/atom relative to the PES from DFT. The error of the bulk FES extracted from WT-MetaD is smaller than $1$meV/f.u. relative to unbiased MD simulations. In the \pto case study, shear strain was not considered but could be included with a straightforward extension within the quadratic theory of electrostriction. We leave for future work the derivation of other terms of the free energy functional, such as polarization gradient corrections ~\cite{wang2017phase, behera2011structure}, with controlled ab initio accuracy.

Our approach can be transferred to other ferroic materials and first-principles atomic models without substantial modification because of the generality of WT-MetaD and the wide applicability of quadratic elastic theory. In the context of atomistic spin dynamics of magnetic crystals~\cite{eriksson2017atomistic} governed by the Landau-Lifshitz Gilbert equations, WT-MetaD could be implemented by projecting the bias force onto the direction that preserves the spin magnitude. For other ferroic materials and models without SO(3) constraints, one could use the same implementation as the one given here for ferroelectrics. Extension to multiferroic materials would involve
dealing with order parameters of higher dimensionality, which may require other sampling methods such as variational enhanced sampling~\cite{invernizzi2017coarse}. To facilitate future studies along these lines, we have developed the software tool ``OpenFerro''~\cite{openferro} which provides automatic differentiation of an on-lattice Hamiltonian and supports enhanced sampling of MD and Landau-Lifshitz-Gilbert simulations. 

  

\textit{Acknowledgement} - 
We thank Linfeng Zhang and Han Wang for their assistance with DeePMD-kit. We thank Han Wang for assistance in implementing the DeepMD Plumed Module. And we thank Michiel Sprik and Karin M. Rabe for fruitful discussions. The majority of the reported work was performed at Princeton University, where P.X., Y.C., W. E, R.C. were supported by the Computational Chemical Sciences Center: Chemistry in Solution and at Interfaces (CSI) funded by U.S. Department of Energy (DOE) Award DE-SC0019394. The work was also performed at Lawrence Berkeley National Laboratory, where P.X. and X.X. was supported by DOE Advanced Scientific Computing Research (ASCR) Applied Mathematics program under Contract No. DE-AC02-05CH11231.
Calculations were performed on the National Energy Research Scientific Computing Center (NERSC), a U.S. Department of Energy Office of Science User Facility operated under Contract No. DE-AC02-05CH11231. Calculations were also performed using the Princeton Research Computing resources at Princeton University, which is a consortium of groups led by the Princeton Institute for Computational Science and Engineering (PICSciE) and Office of Information Technology's Research Computing.

\textit{Data Availability} - 
The datasets, models, and scripts that support the findings of this study are publicly available on GitHub~\cite{alldata}. 

\bibliography{ferro}
\onecolumngrid
\section*{Appendixes}
\twocolumngrid

\textit{Appendix A - 1D free energy profile $G(T, d_c)$}

\begin{figure}[bth]
    \centering
    \includegraphics[width=\linewidth]{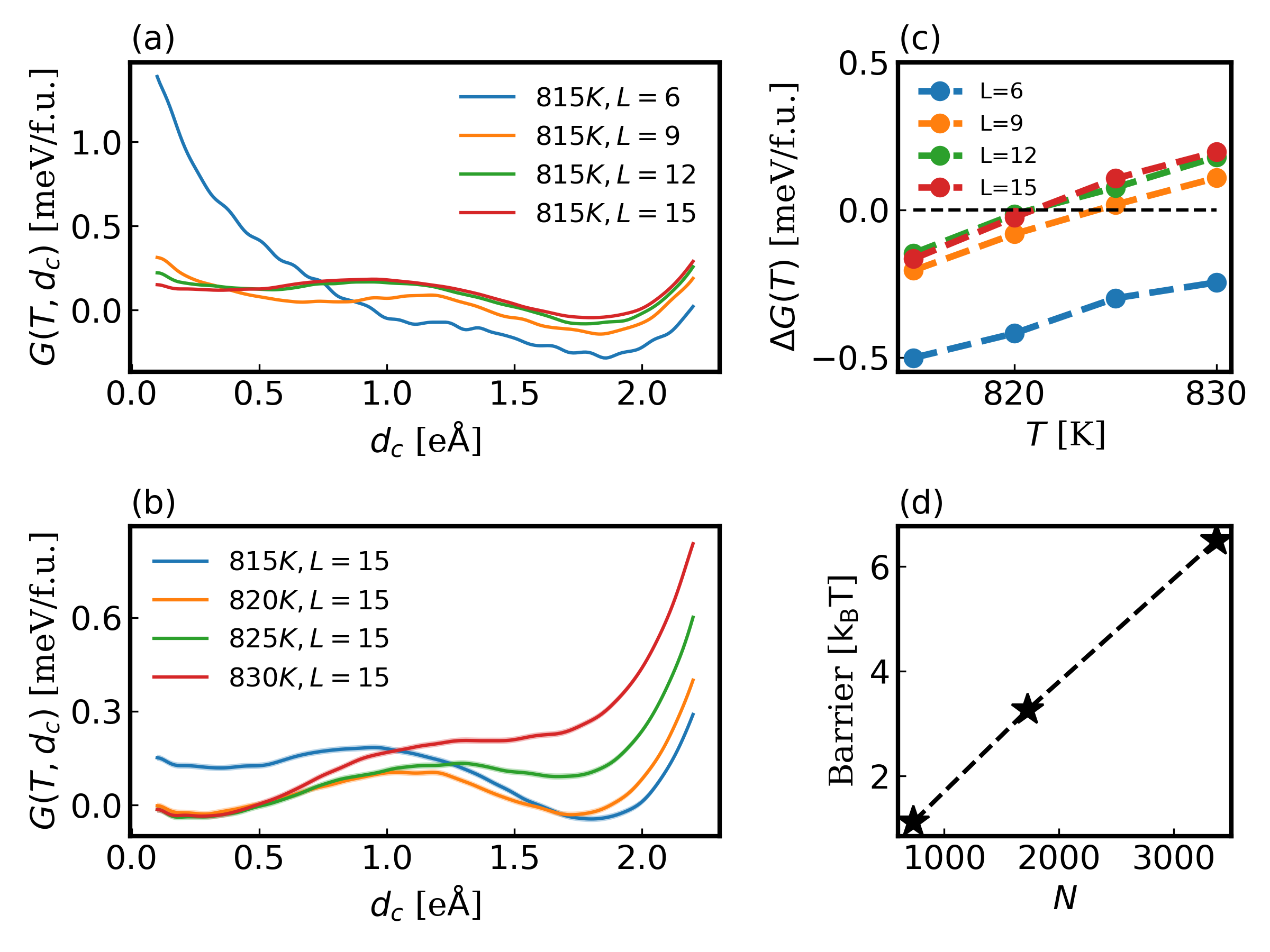}
    \caption{  (a) $G(T=815\text{K},d_c)$ as a function of $d_c$ for different values of $L$. 
    (b) $G(T,d_c)$ as a function of $d_c$ for different temperatures. 
    (c) Free energy difference $\Delta G(T)$ computed with different supercell sizes. (d) Size dependence of the free energy barrier for the transition from the ferroelectric state to the paraelectric state at T=820K. }
    \label{fig:fes1d}
\end{figure}

We perform WT-MetaD to directly compute $G(T, d_c)$ for $d_c\in[0,2.5]$eA using simulation boxes with $L\times L \times L$ elementary cells, for $T\in [815,820,825,830]$K. 
Comparing the results obtained with $L=6,9,12,15$, we find that $G(T, d_c)/L^3$ converges uniformly with increasing $L$. When $L\geq 9$, the finite-size effect on $G(T, d_c)/L^3$ is marginal. However, when $L<9$, the paraelectric phase is substantially destabilized. This is evidenced in Fig.~\ref{fig:fes1d}(a) for $T=815$K, where $L\geq 9$ yields a global free energy minimum near $d_c=2\text{e\AA}$, corresponding to a stable ferroelectric phase, and a local free energy minimum in the vicinity of $d_c=0$, corresponding to a metastable paraelectric phase, which disappears when $L=6$. 
These finite-size effects originate from thermal disordering in the paraelectric phase. Accurate modeling of the entropy gained from disorder requires supercells sufficiently larger than the correlation length ($\xi$) of the cell dipoles. It was found in~\cite{xie2024pto} that $\xi \approx 1$nm, which explains the stark difference between the free energy profiles with $L=6$ and $L=9$. 
We see in Fig.~\ref{fig:fes1d}(a) that the local free energy minimum associated with the paraelectric state is located at a non-zero $d_c$ even for the larger supercells with $L=15$.  This is an artifact from the definition of the 1D free energy profile $G(T, d_c)$. $G(T, d_c)$ must have a logarithmic-like singularity for $d_c\rightarrow 0$, which is consistent with the flat minimum about $\boldsymbol{\mathcal{P}}=0$ in the relevant 3D FES $\mathcal{G} (T, \boldsymbol{\mathcal{P}})$. 
Fig.~\ref{fig:fes1d}(b) shows that the paraelectric phase switches from metastable to stable when $T$ increases from $T=815$K (below $T_c$) to $830$K (above $T_c$). The width of the thermal hysteresis is roughly 15K, which is consistent with the experimental value ($16\pm 3$)K~\cite{Rossetti_2005}. The phenomenological theory of a structural phase transition involving a crystal lattice mismatch~\cite{schwarz1995thermodynamics,jin2019macroscopic} pointed out that such thermal hysteresis is indeed associated with a volume-law free energy barrier, and this hysteresis ~\footnote{A heuristic calculation with Eq. (20c) of Ref.~\cite{jin2019macroscopic} yields a width of thermal hysteresis in the same order of magnitude as the experimental value.} can be observed in experiments regardless of the heating/cooling rate.

\begin{figure*}[hbt]
    \centering
    \includegraphics[width=\linewidth]{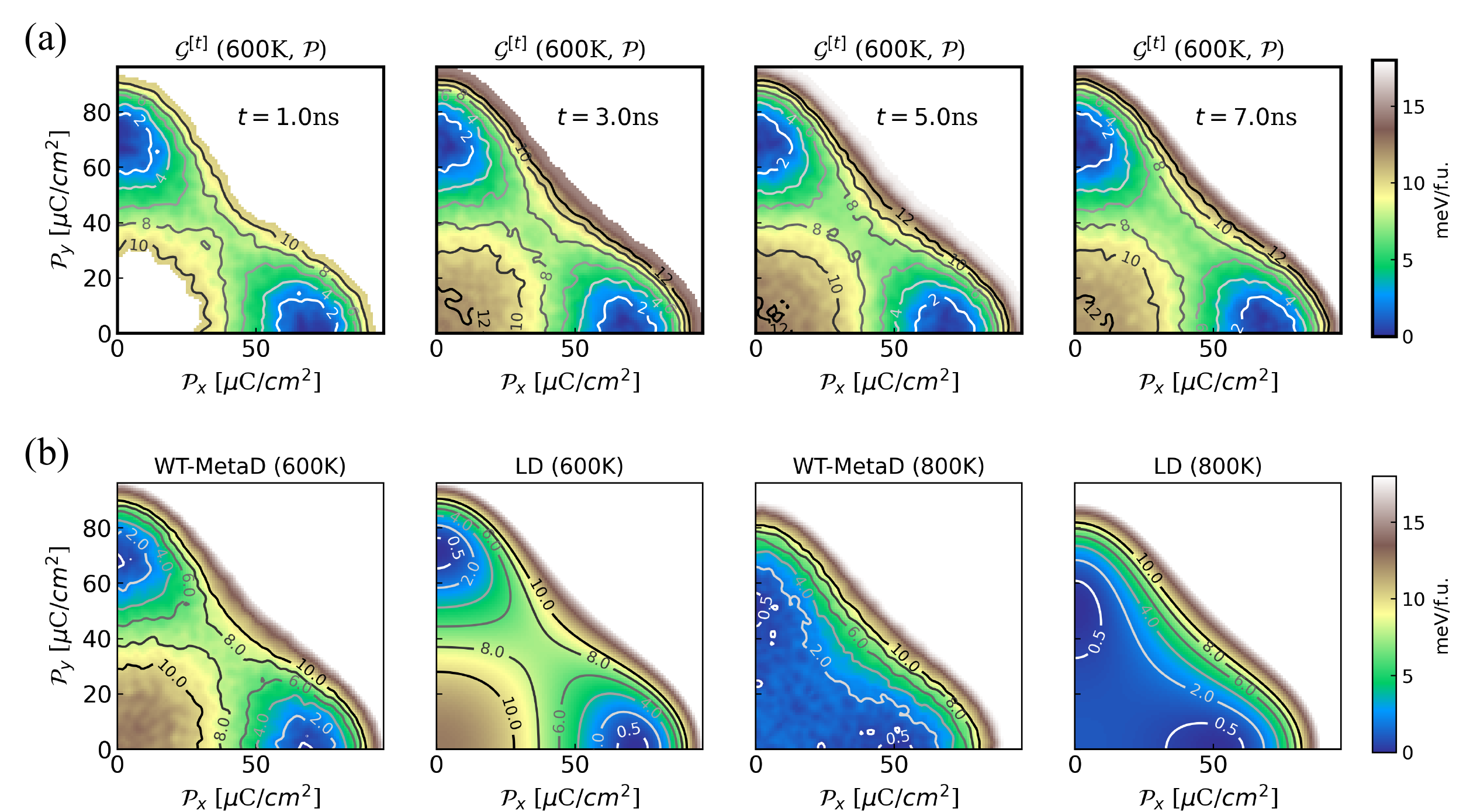}
    \caption{(a) Heat and contour maps of the $\mathcal{P}_z=0$ section of the instantaneous free energy $\mathcal{G}^{[t]}( T, \boldsymbol{\mathcal{P}})$ at different instants of an 8ns-long WT-MetaD simulation. The white areas indicate unexplored regions. (b) The $\mathcal{P}_z=0$ sections of the $\mathcal{G}( T, \boldsymbol{\mathcal{P}})$ and $\mathcal{G}_{\text{LD}}( T, \boldsymbol{\mathcal{P}})$ at two different temperatures. The first and third panels are associated with $\mathcal{G}( T, \boldsymbol{\mathcal{P}})$ from WT-MetaD simulations. The second and fourth panels are associated with the optimal Landau-Devonshire (LD) model $\mathcal{G}_{\text{LD}}( T, \boldsymbol{\mathcal{P}})$. }
    \label{fes2D-converge}
\end{figure*}

WT-MetaD is also useful to compute the absolute free energy difference $\Delta G(T)$ between the ferroelectric and the paraelectric phase, which can be formally defined as
\begin{equation}\label{dG_def}
\small
    \Delta G(T) = -k_BT \ln \frac{\int  d \mathbf{R} e^{-\beta \mathcal{H}(\mathbf{R})}  \Theta(\|\boldsymbol{\mathcal{P}}(\mathbf{R})\|\Omega(\mathbf{R}) - \mathfrak{P}) }{\int  d \mathbf{R} e^{-\beta \mathcal{H}(\mathbf{R})}  (1-\Theta(\|\boldsymbol{\mathcal{P}}(\mathbf{R})\|\Omega(\mathbf{R}) - \mathfrak{P}))}.
\end{equation}
Here, $\Theta$ is the Heaviside step function. $\mathfrak{P}$ is a threshold parameter that distinguishes the two phases. We set $\mathfrak{P}=1 \mathrm{e\AA}$. In the thermodynamic limit, $\Delta G(T)$ will not be affected by small changes in the value of $\mathfrak{P}$. Eq.~\eqref{dG_def} can be transformed into 
\begin{equation}\label{fes_diff}
    \small
    \Delta G(T) = -k_BT\ln \frac{\langle \Theta(\|\boldsymbol{\mathcal{P}}(\mathbf{R})\|\Omega(\mathbf{R})- \mathfrak{P}) e^{\beta {V_b}} \rangle_{V_b} }{\langle (1-\Theta(\|\boldsymbol{\mathcal{P}}(\mathbf{R})\|\Omega(\mathbf{R}) - \mathfrak{P}))e^{\beta {V_b}}  \rangle_{V_b}},
\end{equation}
where $\langle \cdot \rangle_{V_b}$ indicates the NPT average in the biased ensemble (where the biasing potential $V_b$ is added to the potential energy $\mathcal{H}$). It is convenient to use the asymptotic $V_b$ of WT-MetaD, which ensures a good sampling of both phases by facilitating frequent barrier crossings (the reweighting technique used here is general and can be used to study the statistics of other collective variables). The calculated values of $\Delta G(T)$ for $L\in[6,9,12,15]$ and $T\in[815,820,825,830]$K are plotted in Fig.~\ref{fig:fes1d}(c), which shows that finite size effects for $L\geq 9$ lead to an error less than 0.1meV/f.u. in $\Delta G(T)$. Linear interpolation for $L=15$ gives $T_c=(821\pm 1)K$.  An error of the order of 0.1meV/f.u. is insignificant compared to the 1meV/atom error of the DP model. The latter may contribute to an error in $T_c$ of the order of $10K$, as a very rough estimation.

Near the phase transition temperature $T_c$, we find that the free energy barrier associated with the transition of a finite simulation cell from the ferroelectric to the paraelectric state scales as the volume ($\sim L^3$), instead of the surface ($\sim L^2$), of the simulation cell. This scaling law is illustrated in Fig.~\ref{fig:fes1d}(d), which shows the free energy barrier calculated at $T=820$K for finite systems with $N=L^3$ elementary cells.
 
\textit{Appendix B - Details of WT-MetaD simulation}


    The WT-MetaD simulations were carried out using the software packages DeePMD-kit~\cite{wang2018deepmd,zeng2023deepmd}, LAMMPS~\cite{thompson2022lammps}, and PLUMED~\cite{bonomi2019promoting, tribello2014plumed} and an additional package~\cite{yixiao} that implements the polarization model as a differentiable collective variable in PLUMED.
    We used a time step $\Delta t=0.5$fs and periodic boundary condition. The isothermal-isobaric condition was maintained with the MTK method ~\cite{martyna1994constant} using the default parameters in LAMMPS. The chosen values of the biasing factor $\gamma$ in each WT-MetaD simulation are reported as the ``BIASFACTOR'' parameter in the publicly available PLUMED scripts (see Data Availability). The guideline for choosing $\gamma$ in this work was to scale $\gamma$ with the number of atoms in the supercell, in order to effectively reduce the free energy barrier height to the order of $k_BT$ for each supercell size. 

The 1D free energy profiles reported in Appendix A were extracted from WT-MetaD simulations that lasted $4$ ns each. The stochastic error in the free-energy profile was estimated from the standard deviation of the fluctuation of the bias potential, which was evaluated in the last 2 ns of the trajectories when WT-MetaD was well equilibrated and $G(T, d_c)$ was well converged. In this way, the statistical error in $G(T, d_c)$ was estimated to be of the order of $0.1$meV/f.u. 

$\mathcal{G}(T,\boldsymbol{\mathcal{P}})$ was computed from WT-MetaD simulations that lasted $8$ns each. The typical evolution of the instantaneous $ \mathcal{G}^{[t]}(T, \boldsymbol{\mathcal{P}})$ during WT-MetaD 
is illustrated in Fig.~\ref{fes2D-converge}(a), where each contour plot was obtained by rescaling the instantaneous bias potential $V^{[t]}$ at time $t\in[0, 8]$ns. The stochastic error amounts to $0.2\sim 0.3$meV/f.u. The stochastic error here is larger than that in $G(T, d_c)$, despite a longer simulation time, due to the higher dimensionality of the order parameter. 
Fig.~\ref{fes2D-converge}(b) shows comparisons between $\mathcal{G}(T,\boldsymbol{\mathcal{P}})$ and the optimal $\mathcal{G}_{\text{LD}}(T,\boldsymbol{\mathcal{P}})$ at $T=$600K and 800K. 

\newpage
\onecolumngrid
\section*{Supplemental Material}

\subsection{Optimal Landau-Devonshire model}
We describe the protocol that we used for defining the optimal Landau-Devonshire (LD) model that denoises our WT-MetaD data.
 
The sixth-order LD model is
\begin{equation}\label{sm:landau} 
\small
    \begin{split}
    \mathcal{G}_{\text{LD}}(T,\mathcal{P})=G_0(T) &+ \alpha_1(T)\|\boldsymbol{\mathcal{P}}\|^2 
     \\
    &+\alpha_{11}(T)(\mathcal{P}_x^4+\mathcal{P}_y^4+\mathcal{P}_z^4) \\
    &+\alpha_{12}(T)(\mathcal{P}_x^2\mathcal{P}_y^2 
    +\mathcal{P}_x^2\mathcal{P}_z^2 +\mathcal{P}_y^2\mathcal{P}_z^2  )\\
    &+\alpha_{111}(T)(\mathcal{P}_x^6+\mathcal{P}_y^6+\mathcal{P}_z^6)\\
    &+\alpha_{112}(T)[
        \mathcal{P}_x^4(\mathcal{P}_y^2+\mathcal{P}_z^2)
        +\mathcal{P}_y^4(\mathcal{P}_z^2+\mathcal{P}_x^2)
        +\mathcal{P}_z^4(\mathcal{P}_x^2+\mathcal{P}_y^2)] \\
    & +\alpha_{123}\mathcal{P}_x^2\mathcal{P}_y^2\mathcal{P}_z^2.
    \end{split}
\end{equation}

In condensed form this equation can be written as
\begin{equation}\label{sm:landau-condensed}
    \mathcal{G}_{\text{LD}}(T,\mathcal{P})-G_0(T) = \sum_{i=1}^6 \alpha_{[i]}(T) \text{p}_{[i]}(\boldsymbol{\mathcal{P}}),
\end{equation} 
where $\alpha_{[1]}(T), \alpha_{[2]}(T), \cdots, \alpha_{[6]}(T)$ label $\alpha_1(T), \alpha_{11}(T), \cdots, \alpha_{123}(T)$.   $\text{p}_{[i]}(\boldsymbol{\mathcal{P}})$ is the polynomial of ${\mathcal{P}_x,\mathcal{P}_y,\mathcal{P}_z}$ associated with $\alpha_{[i]}(T)$ in Eq.~\eqref{sm:landau}.
        
Assuming that each LD coefficient has a linear temperature dependence, 
$\alpha_{[i]}(T) = \alpha_{[i],0} + \alpha_{[i],1} T$, where $ \alpha_{[i],0}$ and  $\alpha_{[i],1}$ are temperature-independent constants.  Then, Eq.~\eqref{sm:landau-condensed}
can be rewritten as 
\begin{equation}\label{sm:ldp}
\mathcal{G}_{\text{LD}}(T,\mathcal{P}) - G_0(T) = \sum_i \alpha_{[i],0} \text{p}_{[i]}(\boldsymbol{\mathcal{P}}) + \sum_i \alpha_{[i],1} T  \text{p}_{[i]}(\boldsymbol{\mathcal{P}}) 
\end{equation}


The optimization problem for the LD coefficients in Eq.~\eqref{sm:ldp} can be conveniently converted into a linear squares fitting problem. 

Recall that the WT-MetaD data are obtained for $T=600, 700, 800, 820, 900$ and $1000$K, respectively.  For each temperature $T$, the dataset includes $\mathcal{G} (T, \boldsymbol{\mathcal{P}})$ values on a dense grid of $\boldsymbol{\mathcal{P}}=(\mathcal{P}_x, \mathcal{P}_y, \mathcal{P}_z)$. Then, the $\mathcal{G} (T, \boldsymbol{\mathcal{P}})$ data form a matrix $\mathcal{G}_{jk}$, where $j=1,2,\cdots, 6$ indicate the temperatures $T=600, 700, \cdots, 1000$K, and $k=1, \cdots, N_s$ labels the sampled $\boldsymbol{\mathcal{P}}$ points on the grid. 

Substituting $\mathcal{G}_{\text{LD}}(T,\boldsymbol{\mathcal{P}})$ in Eq.~\eqref{sm:ldp} with $\mathcal{G}_{jk}$, $G_0(T)$ with the undetermined constant $G_{0,j}$, and $\text{p}_{[i]}(\boldsymbol{\mathcal{P}})$ with its value $\text{p}_{[i], k}$ directly calculated on the grid, one obtains
\begin{equation}\label{sm:linearfit}
    \mathcal{G}_{jk} = \sum_i \alpha_{[i],0} \text{p}_{[i], k} + \sum_i \alpha_{[i],1} T_j \text{p}_{[i], k} + G_{0,j}
\end{equation}

Eq.~\eqref{sm:linearfit} is a linear relation between $\mathcal{G}_{jk}$ and the optimal $\alpha_{[i],0}$, $\alpha_{[i],1}$, and $G_{0,j}$, which can be solved by standard least square fitting methods. 
A Python implementation of the approach sketched here can be found on Github\footnote{\url{github.com/salinelake/ab_initio_PbTiO3/blob/main/Metadynamics/FES_3D/FES_Landau.ipynb}}. 

\subsection{Optimal electrostriction theory}

Electrostriction theory associates the bulk free energy surface (FES) $\mathcal{F}$ with the polarization-only FES $\mathcal{G}$ through 
\begin{equation}
\mathcal{F}(T,\boldsymbol{\mathcal{P}},\boldsymbol{\eta}) =\mathcal{G}(T,\boldsymbol{\mathcal{P}})  +  \frac{N}{2} (\boldsymbol{\eta}^T-\boldsymbol{Q}^T\hat{Z}^T(T)) \hat{B}(T) (\boldsymbol{\eta} - \hat{Z}(T)\boldsymbol{Q}). 
\end{equation}
$\boldsymbol{Q}^T$ is the row vector $(\mathcal{P}_x^2, \mathcal{P}_y^2,\mathcal{P}_z^2)$.  The $3\times 3$ matrix $\hat{Z}(T)$ has equal diagonal elements ($Z_{11}(T)$) and equal off-diagonal elements ($Z_{12}(T)$). The $3\times 3$ matrix  $\hat{B}$ has equal diagonal elements ($B_{11}(T)$) and equal off-diagonal elements ($B_{12}(T)$). We call $\boldsymbol{\zeta}=\boldsymbol{\eta}-\hat{Z}\boldsymbol{Q}$ the residual vector. 

In the following, we describe the scheme for calculating $l^{\text{ref}}_i(T), Z_{11}(T)$, $Z_{12}(T)$, $B_{11}(T)$, $B_{12}(T)$ from WT-MetaD simulations, at each temperature $T$. The WT-MetaD simulations yield trajectories of the polarization vector $\boldsymbol{\mathcal{P}}$ and the cell parameters $\boldsymbol{l}=(l_x,l_y,l_z)$. The instantaneous strain is associated with the cell parameters by $\eta_i = \frac{l_i-l^{\text{ref}}_i(T)}{l^{\text{ref}}_i(T)}$. 
Our approach involves three steps. 

\subsubsection{Determine $l^{\text{ref}}_i(T), Z_{11}(T)$ and $Z_{12}(T)$}

If $l^{\text{ref}}$ were known, $\boldsymbol{\eta}$ and $\boldsymbol{Q}$ could be directly obtained from the WT-MetaD trajectories. 
Then, one could determine $\hat{Z}(T)$ directly by minimizing the $\mathrm{L^2}$-norm of the residual vector.  
However,  $l^{\text{ref}}_i$ is unknown. So we need to include also the determination of the optimal $l^{\text{ref}}_i$ in the optimization task.

Since $(\boldsymbol{\eta}-\hat{Z}\boldsymbol{Q})_i= \frac{l_i-l^{\text{ref}}_i(T)}{l^{\text{ref}}_i(T)} - \sum_j Z_{ij}(T) \mathcal{P}_j^2=\frac{1}{l^{\text{ref}}_i(T)} (l_i-l^{\text{ref}}_i(T) - \sum_j l^{\text{ref}}_i(T) Z_{ij} \mathcal{P}_j^2)$, we define the objective function as
\begin{equation}
    \mathcal{L}=\sum_{i=x,y,z} \mathcal{L}_i =  \sum_{i=x,y,z}(l_i-l^{\text{ref}}_i(T) - \sum_j l^{\text{ref}}_i(T) Z_{ij}(T) \mathcal{P}_j^2)^2.
\end{equation}

Assuming $l^{\text{ref}}(T)=l^{\text{ref}}_x(T)=l^{\text{ref}}_y(T)=l^{\text{ref}}_z(T)$, we have 
\[L'_x = (l_x-l^{\text{ref}}(T) -  l^{\text{ref}}(T) Z_{11}(T) \mathcal{P}_x^2 - l^{\text{ref}}(T) Z_{12}(T) (\mathcal{P}_y^2+\mathcal{P}_z^2))^2,\]
\[L'_y = (l_y-l^{\text{ref}}(T) -  l^{\text{ref}}(T) Z_{11}(T) \mathcal{P}_y^2 - l^{\text{ref}}(T) Z_{12}(T) (\mathcal{P}_x^2+\mathcal{P}_z^2))^2,\]
and
\[L'_z = (l_z-l^{\text{ref}}(T) -  l^{\text{ref}}(T) Z_{11}(T) \mathcal{P}_z^2 - l^{\text{ref}}(T) Z_{12}(T)(\mathcal{P}_x^2+\mathcal{P}_y^2))^2.\]

These equations represent a least square problem treating $l^{\text{ref}}(T)$, $l^{\text{ref}}(T)Z_{11}(T)$, and $l^{\text{ref}}(T) Z_{12}(T)$ as three unknown variables, which can be solved by the standard least square fitting routines, for each temperature $T$. 

\subsubsection{Determine $B_{11}(T)$, $B_{12}(T)$}

In the previous step, we obtained the optimal $l^{\text{ref}}(T)$,  $Z_{11}(T)$, and $Z_{12}(T)$, for each temperature. Then, we can calculate the residual vector $\boldsymbol{\zeta}=\boldsymbol{\eta}-\hat{Z}\boldsymbol{Q}$ along each WT-MetaD trajectory.  

At each temperature $T$, the so obtained trajectory of $\boldsymbol{\zeta}$ obeys a three-dimensional normal distribution. Then, we calculate the covariance matrix $\hat{\mathcal{C}}(T)$ associated with the three-dimensional normal distribution.  
The matrix $\hat{B}(T)$ is associated with the covariance matrix by $\hat{\mathcal{C}}(T)=(\beta N)^{-1}\hat{B}^{-1}(T)$. $N$ is the total number of elementary cells in the simulation box. 
Therefore, we determine $\hat{B}(T)$ from $\hat{B}(T) = (\beta N) \hat{\mathcal{C}}^{-1}(T)$. 

\subsubsection{Interpolate all parameters as functions of temperature}

Finally, we interpolate $l^{\text{ref}}_i(T), Z_{11}(T)$, $Z_{12}(T)$, $B_{11}(T)$, $B_{12}(T)$, obtained independently for each temperature $T$, as linear or quadratic functions of the temperature. 

A Python implementation of the approach sketched here can be found on Github  \footnote{\url{github.com/salinelake/ab_initio_PbTiO3/blob/main/Metadynamics/FES_6D/electrostriction_biaxial.ipynb}}.

\end{document}